\documentclass[12pt]{article}
\begin{document}

\title{\bf Teleparallel Version of the Stationary Axisymmetric Solutions
and their Energy Contents}

\author{M. Sharif \thanks{msharif@math.pu.edu.pk} and M. Jamil Amir
\thanks{mjamil.dgk@gmail.com}\\
Department of Mathematics, University of the Punjab,\\
Quaid-e-Azam Campus, Lahore-54590, Pakistan.}

\date{}

\maketitle

\begin{abstract}
This work contains the teleparallel version of the stationary
axisymmetric solutions. We obtain the tetrad and the torsion fields
representing these solutions. The tensor, vector and axial-vector
parts of the torsion tensor are evaluated. It is found that the
axial-vector has component only along $\rho$ and $z$ directions. The
three possibilities of the axial vector depending on the metric
function $B$ are discussed. The vector related with spin has also
been evaluated and the corresponding extra Hamiltonian is furnished.
Further, we use the teleparallel version of M$\ddot{o}$ller
prescription to find the energy-momentum distribution of the
solutions. It is interesting to note that (for $\lambda=1$) energy
and momentum densities in teleparallel theory are equal to the
corresponding quantities in GR plus an additional quantity in each,
which may become equal under certain conditions. Finally, we discuss
the two special cases of the stationary axisymmetric solutions.
\end{abstract}


\section{Introduction}

The attempts made by Einstein and his followers to unify gravitation
with other interactions led to the investigation of structures of
gravitation other than the metric tensor. These structures yield the
metric tensor as a by product. Tetrad field is one of these
structures which leads to the theory of teleparallel gravity (TPG)
[1,2]. TPG is an alternative theory of gravity which corresponds to
a gauge theory of translation group [3,4] based on
Weitzenb$\ddot{o}$ck geometry [5]. This theory is characterized by
the vanishing of curvature identically while the torsion is taken to
be non-zero. In TPG, gravitation is attributed to torsion [4] which
plays a role of force [6]. In General Relativity (GR), gravitation
geometrizes the underlying spacetime. The translational gauge
potentials appear as a non-trivial part of the tetrad field and
induce a teleparallel (TP) structure on spacetime which is directly
related to the presence of a gravitational field. In some other
theories [3-8], torsion is only relevant when spins are important
[9]. This point of view indicates that torsion might represent
additional degrees of freedom as compared to curvature. As a result,
some new physics may be associated with it. Teleparallelism is
naturally formulated by gauging external (spacetime) translations
which are closely related to the group of general coordinate
transformations underlying GR. Thus the energy-momentum tensor
represents the matter source in the field equations of tetradic
theories of gravity like in GR.

There is a large literature available [10] about the study of TP
versions of the exact solutions of GR. Recently, Pereira, et al.
[11] obtained the TP versions of the Schwarzschild and the
stationary axisymmetric Kerr solutions of GR. They proved that the
axial-vector torsion plays the role of the gravitomagnetic component
of the gravitational field in the case of slow rotation and weak
field approximations. In a previous paper [12], we have found the TP
versions of the Friedmann models and of the Lewis-Papapetrou
spacetimes, and also discussed their axial-vectors.

There has been a longstanding, controversial and still unresolved
problem of the localization of energy (i.e., to express it as a
unique tensor quantity) in GR [13]. Einstein [14] introduced the
energy-momentum pseudo-tensor and then Landau-Lifshitz [15],
Papapetrou [16], Bergmann [17], Tolman [18] and Weinberg [19]
proposed their own prescriptions to resolve this issue. All these
prescriptions can provide meaningful results only in Cartesian
coordinates. But M$\ddot{o}$ller [20] introduced a
coordinate-independent prescription. The idea of
coordinate-independent quasi-local mass was introduced [21] by
associating a Hamiltonian term to each gravitational energy-momentum
pseudo-tensor. Later, a Hamiltonian approach in the frame of
Schwinger condition [22] was developed, followed by the construction
of a Lagrangian density of TP equivalent to GR [4,6,23,24]. Many
authors explored several examples in the framework of GR and found
that different energy-momentum complexes can give either the same
[25] or different [26] results for a given spacetime.

Mikhail et al. [27] defined the superpotential in the Moller's
tetrad theory which has been used to find the energy in TPG. Vargas
[28] defined the TP version of Bergman, Einstein and Landau-Lifshitz
prescriptions and found that the total energy of the closed
Friedman-Robinson-Walker universe is zero by using the last two
prescriptions. This agrees with the results of GR available in
literature [29]. Later, many authors [30] used TP version of these
prescriptions and showed that energy may be localized in TPG.
Keeping these points in mind, this paper is addressed to the
following two problems: We obtain TP version of the stationary
axisymmetric solutions and then calculate the axial-vector part of
the torsion tensor. The energy-momentum distribution of the
solutions is explored by using the TP version of M$\ddot{o}$ller
prescription.

The scheme adopted in this paper is as follows. In section $2$, we
shall review the basic concepts of TP theory. Section $3$ contains
the TP version of the stationary axisymmetric solutions and the
tensor, vector and axial-vector parts of the torsion tensor. Section
$4$ is devoted to evaluate the energy-momentum distribution for this
family of solutions using the TP version of M$\ddot{o}$ller
prescription. In section $5$, we present two special solutions for
this class of metrics and investigate the corresponding quantities.
The last section contains a summary and a discussion of the results
obtained.

\section{An Overview of the Teleparallel Theory}

In teleparallel theory, the connection is a Weitzenb$\ddot{o}$ck
connection given as [31]
\begin{eqnarray}
{\Gamma^\theta}_{\mu\nu}={{h_a}^\theta}\partial_\nu{h^a}_\mu,
\end{eqnarray}
where ${h_a}^\nu$ is a non-trivial tetrad. Its inverse field is
denoted by ${h^a}_\mu$ and satisfy the relations
\begin{eqnarray}
{h^a}_\mu{h_a}^\nu={\delta_\mu}^\nu; \quad\
{h^a}_\mu{h_b}^\mu={\delta^a}_b.
\end{eqnarray}
In this paper, the Latin alphabet $(a,b,c,...=0,1,2,3)$ will be used
to denote tangent space indices and the Greek alphabet
$(\mu,\nu,\rho,...=0,1,2,3)$ to denote spacetime indices. The
Riemannian metric in TP theory arises as a by product [4] of the
tetrad field given by
\begin{equation}
g_{\mu\nu}=\eta_{ab}{h^a}_\mu{h^b}_\nu,
\end{equation}
where $\eta_{ab}$ is the Minkowski metric
$\eta_{ab}=diag(+1,-1,-1,-1)$. For the Weitzenb$\ddot{o}$ck
spacetime, the torsion is defined as [2]
\begin{equation}
{T^\theta}_{\mu\nu}={\Gamma^\theta}_{\nu\mu}-{\Gamma^\theta}_{\mu\nu}
\end{equation}
which is antisymmetric w.r.t. its last two indices. Due to the
requirement of absolute parallelism, the curvature of the
Weitzenb$\ddot{o}$ck connection vanishes identically. The
Weitzenb$\ddot{o}$ck connection also satisfies the relation
\begin{equation}
{{\Gamma^{0}}^\theta}_{\mu\nu}={\Gamma^\theta}_{\mu\nu}
-{K^{\theta}}_{\mu\nu},
\end{equation}
where
\begin{equation}
{K^\theta}_{\mu\nu}=\frac{1}{2}[{{T_\mu}^\theta}_\nu+{{T_\nu}^
\theta}_\mu-{T^\theta}_{\mu\nu}]
\end{equation}
is the {\bf contortion tensor} and ${{\Gamma^{0}}^\theta}_{\mu\nu} $
are the Christoffel symbols in GR. The torsion tensor of the
Weitzenb$\ddot{o}$ck connection can be decomposed into three
irreducible parts under the group of global Lorentz transformations
[4]: the tensor part
\begin{equation}
t_{\lambda\mu\nu}={\frac{1}{2}}(T_{\lambda\mu\nu}
+T_{\mu\lambda\nu})+{\frac{1}{6}}(g_{\nu\lambda}V_\mu
+g_{\nu\mu}V_\lambda)-{\frac{1}{3}}g_{\lambda\mu}V_\nu,
\end{equation}
the vector part
\begin{equation}
{V_\mu}={T^\nu}_{\nu\mu}
\end{equation}
and the axial-vector part
\begin{equation}
{A^\mu}=\frac{1}{6}\epsilon^{\mu\nu\rho\sigma} T_{\nu\rho\sigma}.
\end{equation}
The torsion tensor can now be expressed in terms of these
irreducible components as follows
\begin{equation}
T_{\lambda\mu\nu}={\frac{1}{2}}(t_{\lambda\mu\nu}-t_{\lambda\nu\mu})
+{\frac{1}{3}}(g_{\lambda\mu}V_\nu -
g_{\lambda\nu}V_\mu)+\epsilon_{\lambda\mu\nu\rho}A^\rho,
\end{equation}
where
\begin{equation}
\epsilon^{\lambda\mu\nu\rho}= \frac{1}{\surd{-g}}
\delta^{\lambda\mu\nu\rho}.
\end{equation}
Here $\delta=\{\delta^{\lambda\mu\nu\rho}\}$ and
$\delta^*=\{\delta_{\lambda\mu\nu\rho}\}$ are completely skew
symmetric tensor densities of weight -1 and +1 respectively [4]. TP
theory provides an alternate description of the Einstein's field
equations which is given by the teleparallel equivalent of GR
[24,31].

Mikhail et al. [27] defined the super-potential (which is
antisymmetric in its last two indices) of the M$\ddot{o}$ller tetrad
theory as
\begin{equation}
{U_\mu}^{\nu\beta}=\frac{\sqrt{-g}}{2\kappa}P_{\chi\rho\sigma}^
{\tau\nu\beta}[{V^\rho}g^{\sigma\chi} g_{\mu\tau}-\lambda
g_{\tau\mu} K^{\chi\rho\sigma}-(1-2\lambda)
g_{\tau\mu}K^{\sigma\rho\chi}],
\end{equation}
where
\begin{equation}
P_{\chi\rho\sigma}^{\tau\nu\beta}= {\delta_\chi}^{\tau}
g_{\rho\sigma}^{\nu\beta}+{\delta_\rho}^{\tau}
g_{\sigma\chi}^{\nu\beta}-{\delta_\sigma}^{\tau}
g_{\chi\rho}^{\nu\beta}
\end{equation}
and $ g_{\rho\sigma}^{\nu\beta}$ is a tensor quantity defined by
\begin{equation}
g_{\rho\sigma}^{\nu\beta}={\delta_\rho}^{\nu}{\delta_\sigma}^{\beta}-
{\delta_\sigma}^{\nu}{\delta_\rho}^{\beta}.
\end{equation}
$K^{\sigma\rho\chi}$ is the contortion tensor given by Eq.(6), $g$
is the determinant of the metric tensor $g_{\mu\nu}$, $\lambda$ is
the free dimensionless coupling constant of TPG, $\kappa$ is the
Einstein constant and $V_\mu$ is the basic vector field given by
Eq.(8). The energy-momentum density is defined as
\begin{equation}
\Xi_\mu^\nu= U_\mu^{\nu\rho},_\rho,
\end{equation}
where comma means ordinary differentiation. The momentum 4-vector
of M$\ddot{o}$ller prescription can be expressed as
\begin{equation}
P_\mu ={\int}_\Sigma {\Xi_\mu^0} dxdydz,
\end{equation}
where  $P_0$ gives the energy and $P_1$, $P_2$ and $P_3$  are the
momentum components while the integration is taken over the
hypersurface element $\Sigma$ described by $x^0=t=constant$. The
energy may be given in the form of surface integral [20] as
\begin{equation}
E=\lim_{r \rightarrow \infty} {\int}_{{r=constant}}
{U_0}^{0\rho}u_\rho dS,
\end{equation}
where $u_\rho$ is the unit three-vector normal to the surface
element $dS$.

\section{Teleparallel Version of the Stationary
Axisymmetric Solutions}

Tupper [32] found five classes of non-null electromagnetic field
plus perfect fluid solutions in which the electromagnetic field does
not inherit one of the symmetries of the spacetime. The metric
representing the stationary axisymmetric solutions is given by [32]
\begin{equation}
ds^2=dt^2-e^{2K}d\rho^2-(F^2-B^2)d\phi^2-e^{2K}dz^2+2Bdtd\phi,
\end{equation}
where $B=B(\rho, z),~K=K(\rho, z)$ and $F=F(\rho)$ are such
functions which satisfy the following relations
\begin{eqnarray}
\dot{B}&=&FW', \quad\ B'=-\frac{1}{4} aF(\dot{W}^2-W'^2), \nonumber\\
K'&=&-\frac{1}{2} aF \dot{W}W', \quad\
\ddot{W}+\dot{F}F^{-1}\dot{W}+W''=0,
\end{eqnarray}
dot and prime denoting the derivatives w.r.t. $\rho$ and $z$
respectively. Here $a$ is constant and $W$ is an arbitrary function
of $\rho$ and $z$, in general. In McIntosh's solution, $W$ is taken
to be $-2bz$ while McLenaghan et. al. solution is obtained by
substituting $W=2\ln\rho$ [33]. The metric given by Eq.(18)
represents five classes of non-null electromagnetic field and
perfect fluid solutions which possesses a metric symmetry not
inherited by the electromagnetic field and admits a homothetic
vector field. Two of these classes contain electrovac solutions as
special cases, while the other three necessarily contain some fluid.
The generalization of this metric is given in [34].

Using the procedure adopted in the papers [11,12], the tetrad
components of the above metric can be written as
\begin{equation}
{h^a}_\mu=\left\lbrack\matrix { 1   &&&   0    &&&   B    &&&   0
\cr 0        &&& e^K\cos\phi &&&   -F\sin\phi    &&&   0 \cr 0 &&&
e^K\sin\phi&&&  F\cos\phi&&& 0 \cr 0        &&&   0    &&& 0 &&&
e^{K} \cr } \right\rbrack
\end{equation}
with its inverse
\begin{equation}
{h_a}^\mu=\left\lbrack\matrix { 1   &&   0    && 0 && 0 \cr
\frac{B}{F}\sin\phi && e^{-K} \cos\phi &&  -\frac{1}{F}\sin\phi &&
0 \cr- \frac{B}{F}\cos\phi  && e^{-K} \sin\phi &&
\frac{1}{F}\cos\phi && 0 \cr 0 && 0 && 0 && e^{-K} \cr }
\right\rbrack.
\end{equation}
The non-vanishing components of the torsion tensor are
\begin{eqnarray}
{T^0}_{12}&=& \dot{B}+\frac{B}{F}(e^K-\dot{F}),\quad\ {T^0}_{32}= B',\nonumber\\
{T^1}_{13}&=&-K', \quad\ {T^2}_{12}=\frac{1}{F}(\dot{F}-e^K),
\quad\ {T^3}_{31}=-\dot{K}.
\end{eqnarray}
Using these expressions in Eqs.(7)-(9), we obtain the following
non-zero components of the tensor part
\begin{eqnarray}
t_{001}&=&\frac{1}{3}[\dot{K}+\frac{1}{F}(\dot{F}-e^K)],\quad\
t_{003}=\frac{1}{3}K', \nonumber\\
t_{010}&=&\frac{1}{6}\{\frac{1}{F}(e^K-\dot{F})-\dot{K}\}=t_{100},\quad\
t_{030}=-\frac{1}{6}K'=t_{300}, \nonumber\\
t_{012}&=&\frac{1}{2}\dot{B}+\frac{B}{6}\{\frac{1}{F}(e^K-\dot{F})
-\dot{K}\}=t_{102}, \nonumber\\
t_{021}&=&-\frac{1}{2}\dot{B}-\frac{B}{3}\{\frac{1}{F}(e^K-\dot{F})
-\dot{K}\}=t_{201}, \nonumber\\
t_{023}&=&-\frac{1}{2}B'+\frac{1}{3}BK'=t_{203},
\quad\ t_{032}={\frac{1}{2}}B'-\frac{1}{6}BK'=t_{302},\nonumber\\
t_{122}&=&\frac{1}{2}\{F(e^K-\dot{F})+B\dot{B}\}+\frac{1}{6}(B^2-F^2)
\{\frac{1}{F}(e^K-\dot{F})-\dot{K}\}=t_{212}, \nonumber\\
t_{120}&=&\frac{B}{6}\{\frac{1}{F}(e^K-\dot{F})-\dot{K}\}=t_{210}, \nonumber\\
t_{113}&=&\frac{2K'}{3}e^{2K},\quad\
t_{131}=-\frac{K'}{3}e^{2K}=t_{311}, \nonumber\\
t_{133}&=&-\frac{e^{2K}}{6}\{\frac{1}{F}(e^K-\dot{F})+2\dot{K}\}=t_{313}, \nonumber\\
t_{221}&=&-F(e^K-\dot{F})-B\dot{B}-\frac{1}{3}(B^2-F^2)\{\frac{1}{F}(e^K-\dot{F})
-\dot{K}\},\nonumber\\
t_{223}&=&-BB'+\frac{K'}{3}(B^2-F^2),\quad\ t_{331}=
\frac{e^{2K}}{3}\{\frac{1}{F}(e^K-\dot{F})+2\dot{K}\},\nonumber\\
t_{322}&=& \frac{1}{2}BB'- \frac{K'}{6}(B^2-F^2)=t_{232}, \quad\
t_{320}=-\frac{1}{6}BK'=t_{232},
\end{eqnarray}
the vector part
\begin{eqnarray}
V_1&=&-\frac{1}{F}(\dot{F}-e^K)-\dot{K},\\
V_3&=&-K',
\end{eqnarray}
and the axial-vector part
\begin{eqnarray}
A^1&=&\frac{B'}{3F}e^{-2K},\\
A^3&=&\frac{\dot{B}}{3F}e^{-2K},
\end{eqnarray}
respectively. The axial-vector component along the $\phi$-direction
vanishes and hence the spacelike axial-vector can be written as
\begin{equation}
\textbf{A}=\sqrt{-g_{11}} A^1 \hat{e}_\rho+ \sqrt{-g_{33}}A^3
\hat{e}_z,
\end{equation}
where $\hat{e}_\rho$ and $\hat{e}_z$ are unit vectors along the
radial and $z$-directions respectively. Substituting $A^1$, $A^3$,
$g_{11}$ and $g_{33}$ in Eq.(28), it follows that
\begin{equation}
\textbf{A}=\frac{e^{-K}}{3F}(B' \hat{e}_\rho+ \dot{B}\hat{e}_z).
\end{equation}
This shows that the axial-vector lies along radial direction if
$B=B(z)$, along $z$-direction if $B=B(\rho)$ and vanishes
identically if $B$ is constant. As the axial-vector torsion
represents the deviation of axial symmetry from cylindrical
symmetry, the symmetry of the underlying spacetime will not be
affected even for $B$ constant. Also, the torsion plays the role
of the gravitational force in TP theory, hence a spinless particle
will obey the force equation [11,24]
\begin{equation}
\frac{du_\rho}{ds}-\Gamma_{\mu\rho\nu} u^\mu u^\nu =T_{\mu\rho\nu}
u^\mu u^\nu.
\end{equation}
The left hand side of this equation is the Weitzenb$\ddot{o}$ck
covariant derivative of $u_\rho$ along the particle world-line.
The appearance of the torsion tensor on its right hand side
indicates that the torsion plays the role of an external force in
TPG. It has been shown, both in GR and TP theories, by many
authors [4,35] that the spin precession of a Dirac particle in
torsion gravity is related to the torsion axial-vector by
\begin{equation}
\frac{d\textbf{S}}{dt}=- \textbf{b}\times \textbf{S},
\end{equation}
where $\textbf{S}$ is the spin vector of a Dirac particle and
$\textbf{b}=\frac{3}{2}\textbf{A}$, with $\textbf{A}$ the spacelike
part of the torsion axial-vector. Thus
\begin{equation}
\textbf{b}= \frac{e^{-K}}{2F}\{B' \hat{e}_\rho+
\dot{B}\hat{e}_z\}.
\end{equation}
The corresponding extra Hamiltonian [36] is given by
\begin{equation}
\delta H= -\textbf{b}.\sigma,
\end{equation}
where $\sigma$ is the spin of the particle [35]. Using Eq.(32), this
takes the form
\begin{equation}
\delta H= -\frac{e^{-K}}{2F}(B' \hat{e}_\rho+
\dot{B}\hat{e}_z).\sigma.
\end{equation}

\section{Teleparallel Energy of the Stationary Axisymmetric Solutions}

In this section we evaluate the component of energy-momentum
densities by using the teleparallel version of M$\ddot{o}$ller
prescription. Multiplying Eqs.(24) and (25) by $g^{11}$ and $g^{33}$
respectively, it follows that
\begin{eqnarray}
V^1&=&\dot{K}e^{-2K}+ \frac{e^{-2K}}{F}(\dot{F}-e^K),\\
V^3&=&K'e^{-2K}.
\end{eqnarray}
In view of Eqs.(6) and (22), the non-vanishing components of the
contorsion tensor are
\begin{eqnarray}
K^{100}&=&-e^{-2K}\{\frac{B^2}{F^3}(e^K-\dot{F})+\frac{B\dot{B}}{F^2}\}
=-K^{010},\nonumber\\
K^{300}&=&-\frac{BB'}{F^2}e^{-2K}=-K^{030}, \quad\
K^{122}=-\frac{e^{-2K}}{F^3}
(e^K-\dot{F})=-K^{212},\nonumber\\
K^{133}&=&\dot{K}e^{-4K}=-K^{313}, \quad\
K^{311}=K'e^{-4K}=-K^{131},  \nonumber\\
K^{102}&=&
K^{120}=e^{-2K}\{\frac{B}{F^3}(e^K-\dot{F})+\frac{\dot{B}}{2F^2}\}
=-K^{012}=-K^{210}, \nonumber\\
K^{302}&=&K^{320}=K^{023}=\frac{B'}{2F^2}
e^{-2K}=-K^{032}=-K^{230}=-K^{203}, \nonumber\\
K^{021}&=&\frac{\dot{B}}{2F^2}e^{-2K}=-K^{201}.
\end{eqnarray}
It should be mentioned here that the contorsion tensor is
antisymmetric w.r.t. its first two indices. Making use of
Eqs.(35)-(37) in Eq.(12), we obtain the required independent
non-vanishing components of the supperpotential in M$\ddot{o}$ller's
tetrad theory as
\begin{eqnarray}
U_0^{01}&=&\frac{1}{\kappa}[e^K-\dot{F}-F \dot{K}+
\frac{1}{2}(1+\lambda) \frac{B\dot{B}}{F}]=-U_0^{10},\nonumber\\
U_0^{03}&=&\frac{1}{\kappa}[-FK'+
\frac{1}{2}(1+\lambda) \frac{BB'}{F}]=-U_0^{30},\nonumber\\
U_0^{21}&=&-\frac{1}{2\kappa}(1+\lambda)\frac{\dot{B}}{F}=-U_0^{12},
\quad\
U_0^{23}=-\frac{1}{2\kappa}(1+\lambda)\frac{B'}{F}=-U_0^{32}, \nonumber\\
U_2^{01}&=&
\frac{1}{\kappa}[B(e^K-\dot{F})+\frac{1}{2}(1+\lambda)\frac{B^2\dot{B}}{F}
+\frac{1}{2}(1-\lambda)\dot{B}F]=-U_2^{10}, \nonumber\\
U_2^{03}&=&\frac{1}{\kappa}[\frac{1}{2}(1+\lambda) \frac{B^2
B'}{F} +\frac{1}{2}(1-\lambda)B'F]=-U_2^{30},\nonumber\\
U_1^{02}&=&\frac{1}{2\kappa F}(\lambda-1)\dot{B}e^{2K}=-U_1^{20}, \nonumber\\
U_3^{02}&=&\frac{1}{2\kappa F}(\lambda-1)B'e^{2K}=-U_1^{30}.
\end{eqnarray}
It is worth mentioning here that the supperpotential is skew
symmetric w.r.t. its last two indices. When we make use of Eqs.(15),
(37), (38) and take $\lambda=1$, the energy density turns out to be
\begin{eqnarray}
\Xi_0^0&=&\frac{1}{\kappa}[\dot{K}e^K-\ddot{F}-\dot{F}\dot{K}-F(\ddot{K}+K'')
+\frac{1}{F^2}\{BF(\ddot{B}+E'') \nonumber\\
&+&({\dot{B}}^2+{B'}^2)F-B \dot{B}\dot{F}\}].
\end{eqnarray}
This implies that
\begin{equation}
{E^d}_{TPT}={E^d}_{GR}
+\frac{1}{\kappa}[\dot{K}e^K-\ddot{F}-\dot{F}\dot{K}-F(\ddot{K}+K'')],
\end{equation}
where $E^d$ stands for energy density. The only non-zero component
of momentum density is along $\phi$-direction and (for $\lambda=1$)
it takes the form
\begin{eqnarray}
\Xi_2^0&=&\frac{1}{\kappa
F^2}\{F^3(\ddot{B}+B'')+B^2F(\ddot{B}+B'')+
2BF({\dot{B}}^2+{B'}^2)-\dot{B}\dot{F}(B^2+F^2)\} \nonumber\\
&+& \frac{1}{\kappa}\{\dot{B}e^K +
B(\dot{K}e^K-\ddot{F})-F(\ddot{B}+B'')\},
\end{eqnarray}
that is,
\begin{equation}
{M^d}_{TPT}={M^d}_{GR} +\frac{1}{\kappa}\{\dot{B}e^K +
B(\dot{K}e^K-\ddot{F})-F(\ddot{B}+B'')\},
\end{equation}
where $M^d$ stands for momentum density.

\section{Special Solutions of the Non-Null Einstein Maxwell Solutions}

In this section, we evaluate the above quantities for some special
cases of the non-null Einstein Maxwell solutions.

\subsection{Electromagnetic Generalization of the G$\ddot{o}$del
Solution}

A special case of the non-null Einstein-Maxwell solutions can be
obtained  by choosing
\begin{equation}
B=\frac{m}{n}e^{n\rho}, \quad\ F=e^{n\rho}, \quad\ K=0.
\end{equation}
This is known as electromagnetic generalization of the
G$\ddot{o}$del solution [32]. When we make use of Eq.(43) in
Eqs.(23)-(27), (29), (32), (34) and (39)-(42), the corresponding
results reduce to
\begin{eqnarray}
t_{001}&=&\frac{1}{3}(n-e^{-n\rho}),\quad\
t_{010}=\frac{1}{6}(e^{-n\rho}-n)=t_{100},\nonumber\\
t_{012}&=&\frac{m}{6n}(1+2ne^{n \rho})=t_{102}, \quad\
t_{021}=-\frac{m}{3n}(1+2ne^{n \rho})=t_{201}, \nonumber\\
t_{122}&=&\frac{e^{n
\rho}}{6n^2}\{m^2+2n^2+2n(m^2-n^2)e^{n\rho}\}=
t_{212}, \nonumber\\
t_{120}&=& \frac{m}{6n}(1-ne^{n \rho})=t_{210},\quad\
t_{133}=\frac{1}{6}(n-e^{-n \rho})=t_{313}, \nonumber\\
t_{221}&=&-\frac{e^{n
\rho}}{3n^2}\{m^2+2n^2+2n(m^2-n^2)e^{n\rho}\}, \nonumber\\
t_{331}&=&-\frac{1}{3}(n-e^{-n \rho}), \\
V_1&=&e^{-n\rho}-n,\quad V_3=0,\\
A^1&=&0,\quad A^3=\frac{m}{3},\\
\textbf{A}&=& \frac{m}{3}\hat{e}_z,\quad \textbf{b}=\frac{m}{2}
\hat{e}_z,\\
\delta H&=&
\frac{m}{2} \hat{e}_z. \sigma,\\
\Xi_0^0&=&\frac{1}{\kappa}(m^2-n^2)e^{n\rho},\\
{E^d}_{TPT}&=& {E^d}_{GR}- \frac{n^2}{\kappa}e^{n\rho} \\
\Xi_2^0&=&\frac{1}{\kappa}(\frac{2m^3}{n})+\frac{m}{\kappa}(1-2ne^{n\rho})e^{n\rho},\\
{M^d}_{TPT}&=&{M^d}_{GR}+\frac{m}{\kappa}(1-2ne^{n\rho})e^{n\rho}.
\end{eqnarray}
The metric (43) reduces to the usual perfect fluid solution when
$m=\sqrt{2}n$ [32], i.e., $B=\sqrt{2}e^{n\rho}$. The corresponding
energy and momentum densities take the form as
\begin{eqnarray}
{E^d}_{TPT}&=& {E^d}_{GR}- \frac{n^2}{\kappa}e^{n\rho} \\
{M^d}_{TPT}&=&{M^d}_{GR}+\frac{\sqrt{2}n}{\kappa}(1-2ne^{n\rho})e^{n\rho}.
\end{eqnarray}

\subsection{The G$\ddot{o}$del Metric}

When we choose $ B=e^{a\rho}, F=\frac{e^{a\rho}}{\sqrt{2}}$  and
$K=0$, the metric given by Eq.(18) reduces to the G$\ddot{o}$del
metric [32]. The results corresponding to Eqs.(23)-(27), (29), (32),
(34) and (39)-(42) take the following form
\begin{eqnarray}
t_{001}&=&\frac{1}{3}(a-\sqrt{2}e^{-a\rho}),\quad\
t_{010}=-\frac{1}{6}(a-\sqrt{2}e^{-a\rho})=t_{100},\nonumber\\
t_{012}&=&\frac{1}{6}(\sqrt{2}+2a e^{a\rho})=t_{102},\quad\
t_{021}=-\frac{1}{6}(2\sqrt{2}+a e^{a\rho})=t_{102}, \nonumber\\
t_{122}&=&\frac{e^{a\rho}}{6}( 2\sqrt{2}+ae^{a\rho})=t_{212},
\quad\ t_{120}=\frac{1}{6}(\sqrt{2}-ae^{a\rho})=t_{210}, \nonumber\\
t_{133}&=&\frac{1}{6}(a-\sqrt{2}e^{-a\rho})=t_{313}, \quad\
t_{221}=-\frac{e^{a\rho}}{3}( 2\sqrt{2}+ae^{a\rho}),\\
t_{331}&=&\frac{1}{3}(a-\sqrt{2}e^{-a\rho}),\\
V_1&=&\sqrt{2}e^{-a\rho}-a,\quad V_3= 0,\\
A^1&=&0,\quad A^3=\frac{\sqrt{2}a}{3},\\
\textbf{A}&=&\frac{\sqrt{2}a}{3} \hat{e}_z,\quad \textbf{b}=
\frac{a}{\sqrt{2}} \hat{e}_z, \\
\delta H&=&\frac{a}{\sqrt{2}} \hat{e}_z. \sigma,\\
\Xi_0^0&=&\frac{\sqrt{2}}{\kappa}a^2
e^{a\rho}-\frac{a^2}{\kappa\sqrt{2}}e^{a\rho},\\
{E^d}_{TPT}&=&{E^d}_{GR}-\frac{a^2}{\kappa\sqrt{2}}e^{a\rho}\\
\Xi_2^0&=&\frac{a^2}{\kappa
\sqrt{2}}e^{2a\rho}+\frac{a}{\kappa}(1-\sqrt{2}ae^{a\rho})e^{a\rho},\\
{M^d}_{TPT}&=&{M^d}_{GR}+\frac{a}{\kappa}(1-\sqrt{2}ae^{a\rho})e^{a\rho}.
\end{eqnarray}

\section{Summary and Discussion}

The purpose of this paper is twofold: Firstly, we have found the TP
version of the non-null Einstein Maxwell solutions. This provides
some interesting features about the axial vector and the
corresponding quantities. Secondly, we have used the TP version of
M$\ddot{o}$ller prescription to evaluate the energy-momentum
distribution of the solutions. The axial-vector torsion of these
solutions has been evaluated. The only non-vanishing components of
the vector part are along the radial and the $z$-directions due to
the cross term $dx^0dx^2$ involving in the metric. This corresponds
to the case of Kerr metric [11], which involves the cross term
$dx^0dx^3$. We also find the vector \textbf{b} which is related to
the spin vector [4] as given by Eq.(32). The axial-vector torsion
lies in the ${\rho}z$-plane, as its component along the
$\phi$-direction vanishes everywhere. The non-inertial force on the
Dirac particle can be represented as a rotation induced torsion of
spacetime.

There arise three possibilities for the axial-vector, depending upon
the metric function $B(\rho, z)$. When $B$ is a function of $z$
only, the axial-vector lies only along the radial direction. When
$B$ is a function of $\rho$ only, the axial-vector will lie along
$z$-direction. The axial-vector vanishes identically for $B$ to be
constant. As the axial-vector represents the deviation from the
symmetry of the underlying spacetime corresponding to an inertial
field with respect to the Dirac particle, the symmetry of the
spacetime will not be affected in the third possibility.
Consequently there exists no inertial field with respect to the
Dirac particle and the spin vector of the Dirac particle becomes
constant. The corresponding extra Hamiltonian is expressed in terms
of the vector \textbf{b} which vanishes when the metric function $B$
is constant, i.e., when the axial-vector becomes zero.

The energy-momentum distribution of the non-null Einstein-Maxwell
solutions has been explored by using the TP version of
M$\ddot{o}$ller prescription. It is found that energy in the TP
theory is equal to the energy in GR (as found by Sharif and Fatima
[37]) plus some additional part. If, for a particular case, we have
$\dot{K}=0$ and $K',~\dot{F}=constant$ (or if $\dot{F},~\dot{K}=0$
and $K'=constant$), then
\begin{equation}
{E^d}_{TPG}={E^d}_{GR}.
\end{equation}
On the other hand, the only non-vanishing component of the momentum
density lies along $\phi$-direction, similar to the case of Kerr
metric [11], due to the cross term appearing in both the metrics.
When we choose $\lambda=1$, it becomes equal to be the momentum in
GR [37] plus an additional quantity. If
$\ddot{F},~\dot{B},~B'',~\dot{K}$ all vanish, then
\begin{equation}
{M^d}_{TPG}={M^d}_{GR}.
\end{equation}

By taking particular values of $E$, $F$ and $K$, we obtain the
electromagnetic generalization of the G$\ddot{o}$del solution and
the G$\ddot{o}$del metric as two special cases. The corresponding
results for both the special cases are obtained. It is shown that,
for the electromagnetic generalization of the G$\ddot{o}$del
solution, Eq.(65) does not hold, while Eq.(66) holds when $m=0$.
However, for the perfect fluid case, i.e., when $m=\sqrt{2}n$, both
Eqs.(65) and (66) hold by taking $n=0$. In the case of the
G$\ddot{o}$del metric, these equations hold if we choose the
arbitrary constant $a=0$. For the special solutions, the vector part
lies along the radial direction while the axial-vector part along
$z$-direction.

We would like to re-iterate here that the tetrad formalism itself
has some advantages which comes mainly from its independence from
the equivalence principle and consequent suitability to the
discussion of quantum issues. In TPG, an energy-momentum gauge
current ${j_i}^\mu$ for the gravitational field can be defined. This
is covariant under a spacetime general coordinate transformation and
transforms covariantly under a global tangent space Lorentz
transformation [38]. It, then, follows that ${j_i}^\mu$ is a true
spacetime tensor but not a tangent space tensor. When we re-write
the gauge field equations in a purely spacetime form, they lead to
the Einstein field equations and the gauge current ${j_i}^\mu$
reduces to the canonical energy-momentum pseudo-tensor of the
gravitational field. Thus TPG seems to provide a more appropriate
environment to deal with the energy problem.

Finally, it is pointed out that we are not claiming that this paper
has resolved the problems of GR using the TPG. This is an attempt to
touch some issues in TPG with the hope that this alternative may
provide more feasible results. Also, it is always an interesting and
enriching to look at things from another point of view. This
endeavor is in itself commendable.

\vspace{0.5cm}


{\bf Acknowledgment}

\vspace{0.5cm}

We would like to thank the Higher Education Commission Islamabad,
Pakistan for its financial support through the {\it Indigenous PhD
5000 Fellowship Program Batch-I} during this work.
\vspace{0.5cm}

{\bf References}

\begin{description}

\item{[1]} M$\ddot{u}$ller-Hoisson, F. and Nitsch, J.: Phys. Rev. {\bf D28}
           (1983)718.

\item{[2]} De Andrade, V. C. and Pereira, J.G.: Gen.Rel.Grav. {\bf 30}(1998)263.

\item{[3]} Hehl, F.W., McCrea, J.D., Mielke, E.W. and Ne'emann, Y.: Phys.
           Rep. {\bf 258}(1995)1.

\item{[4]} Hayashi, K. and Tshirafuji : Phys. Rev. {\bf D19}(1979)3524.

\item{[5]} Weitzenb$\ddot{o}$ck, R.: {\it Invarianten Theorie}
           (Gronningen: Noordhoft, 1923).

\item{[6]} De Andrade, V.C. and Pereira,  J.G.: Phys. Rev. {\bf
           D56}(1997)4689.

\item{[7]} Gronwald, F. and Hehl, F.W.: {\it Proceedings of the
           School of Cosmology and Gravitation on Quantum Gravity}, Eric,
           Italy ed. Bergmann, P.G. et al. (World Scientific, 1995);\\
           Blagojecvic, M. {\it Gravitation and Gauge Symmetries} (IOP
           publishing, 2002).

\item{[8]} Hammond, R.T.: Rep. Prog. Phys. {\bf 65}(2002)599.

\item{[9]} Gronwald, F. and Hehl, F.W.: {\it On the Gauge Aspects of Gravity,
           Proceedings of the 14th School of Cosmology and Gravitation},
           Eric, Italy ed. Bergmann, P.G. et al. (World Scientific, 1996).

\item{[10]} Hehl, F.W. and Macias, A.: Int. J. Mod. Phys. {\bf
D8}(1999)399;\\
Obukhov, Yu N., Vlachynsky, E.J., Esser, W., Tresguerres, R. and
Hehl, F.W.: Phys. Lett. {\bf A220}(1996)1;\\
Baekler, P., Gurses, M., Hehl, F.W. and McCrea, J.D.: Phys. Lett.
{\bf A128}(1988)245;\\
Vlachynsky, E.J. Esser, W., Tresguerres, R. and Hehl, F.W.: Class.
Quant. Grav. {\bf 13}(1996)3253;\\
Ho, J.K., Chern, D.C. and Nester, J.M.: Chin. J. Phys. {\bf
35}(1997)640;\\
Hehl, F.W., Lord, E.A. and Smally, L.L.: Gen. Rel. Grav. {\bf 13}
(1981)1037;\\
Kawa, T. and Toma, N.: Prog. Theor. Phys. {\bf 87}(1992)583;\\
Nashed, G.G.L.: Phys. Rev. \textbf{D66}(2002)060415; Gen. Rel. Grav.
\textbf{34}(2002)1074.

\item{[11]} Pereira, J.G., Vargas, T. and Zhang, C.M.: Class. Quantum Grav.
            {\bf 18}(2001)833.

\item{[12]} Sharif, M. and Amir, M.J.: Gen. Rel. Grav. \textbf{38}(2006)1735.

\item{[13]} Misner, C.W., Thorne, K.S. and Wheeler, J.A.: \textit{Gravitation}
            (Freeman, New York, 1973).

\item{[14]} Einstein, A.: Sitzungsber. Preus. Akad. Wiss. Berlin (Math. Phys.)
            778(1915), Addendum ibid 779(1915).

\item{[15]} Landau, L.D. and Lifshitz, E.M.: \textit{The Classical Theory
            of Fields} (Addison-Wesley Press, New York, 1962).

\item{[16]} Papapetrou, A.: \textit{Proc. R. Irish Acad. } \textbf{A52}(1948)11.

\item{[17]} Bergman, P.G. and Thomson, R.: Phys. Rev.
            \textbf{89}(1958)400.

\item{[18]} Tolman, R.C.: \textit{Relativity, Thermodynamics and
           Cosmology} (Oxford University Press, Oxford, 1934).

\item{[19]} Weinberg, S.: \textit{Gravitation and Cosmology} (Wiley, New
           York, 1972).

\item{[20]} M$\ddot{o}$ller, C.: Ann. Phys. (N.Y.) \textbf{4}(1958)347.

\item{[21]} Chang, C.C. and Nester, J.M.: Phys. Rev. Lett. \textbf{83}
            (1999)1897 and references therein.

\item{[22]} Schwinger, J.: Phys. Rev. \textbf{130}(1963)1253.

\item{[23]} De Andrade, V.L, Guillen, L.C.T and Pereira, J.G.: Phys. Rev. Lett.
            {\bf 84}(2000)4533.

\item{[24]} Aldrovendi, R. and Pereira, J.G.: {\it An Introduction to
            Gravitation Theory} (preprint).

\item{[25]} Virbhadra, K.S.: Phys. Rev. \textbf{D60}(1999)104041;
\textit{ibid} \textbf{D42}(1990)2919; Phys. Lett.
\textbf{B331}(1994)302;\\
Virbhadra, K.S. and Parikh, J.C.: Phys. Lett.
\textbf{b317}(1993)312;\\
Rosen, N. and Virbhadra, K.S.: Gen. Rel. Grav.
\textbf{25}(1993)429;\\
Xulu, S.S.: Astrophys. Space Sci. \textbf{283}(2003)23.

\item{[26]} Sharif, M.: Int. J. Mod. Phys. \textbf{A17}(2002)1175;
\textit{ibid} \textbf{A18}(2003)4361; \textbf{A19}(2004)1495;
\textbf{D13}(2004)1019;\\
Sharif, M. and Fatima, T.: Nouvo Cim.
\textbf{B120}(2005)533.

\item{[27]} Mikhail, F.I., Wanas, M.I., Hindawi, A. and Lashin, E.I.: Int. J. Theo.
            Phys. \textbf{32}(1993)1627.

\item{[28]} Vargas, T.: Gen. Rel. Grav. \textbf{36}(2004)1255.

\item{[29]} Penrose, R.: \textit{Proc. Roy. Soc., London
}\textbf{A381}(1982)53;\\
Tod, K.P.: \textit{Proc. Roy. Soc., London }\textbf{A388}(1983)457.

\item{[30]} Nashed, G.G.L.: Nuovo Cim. \textbf{B119}(2004)967;\\
Salti, M., Havare, A.: Int. J. of Mod. Phys.
\textbf{A20}(2005)2169;\\
Salti, M.: Int. J. of Mod. Phys. \textbf{A20}(2005)2175; Space Sci.
\textbf{229}(2005)159;\\
Aydogdu, O. and Salti, M.: Astrophys. Space Sci.
\textbf{229}(2005)227;\\
Aydogdu, O., Salti, M. and Korunur, M.: Acta Phys. Slov.
\textbf{55}(2005)537;\\
Sharif, M. and Amir, M.J.: Mod. Phys. Lett. \textbf{A22}(2007)425;\\
Sezgin, A., Melis, A. and Tarhan, I.: Acta Physica Polonica
\textbf{B} (to appaer).

\item{[31]} Aldrovandi and Pereira, J.G.: {\it An Introduction to
            Geometrical Physics} (World Scientific, 1995).

\item{[32]} Tupper, B.O.J.: Class. Quantum Grav. \textbf{1}(1984)71.

\item{[33]} Tupper, B.O.J.: Class. Quantum Grav. \textbf{2}(1985)427.

\item{[34]} Stephani, H. Kramer, D., MacCallum, M.A.H., Heonselaers, C. and
            Hearlt, E.: \textit{Exact Solutions of Einstein's Field Equations}
            (Cambridge University Press, 2003).

\item{[35]} Mashhoon, B.: Class. Quantum Grav. \textbf{17}(2000)2399.

\item{[36]} Zhang, C.M. and Beesham, A.: Mod. Phys. Lett. \textbf{A16}(2001)2319.

\item{[37]} Sharif, M. and Fatima, T.: Int. J. Mod. Phys. \textbf{A20}(2005)4309.

\item{[38]} De Andrade, V.C., Arcos, H.I. and Pereira,  J.G.: PoS WC 2004 (2004)028.

\end{description}
\end{document}